\documentclass[chicago,numberedappendix,tighten]{emulateapj}
\usepackage{amsmath}
\usepackage{amssymb}
\usepackage{footnote}
\usepackage[normalem]{ulem}
\usepackage[dvips]{color}
\begin{document}
\title{Inferring the composition of super-Jupiter mass companions of 
pulsars\\
 with radio line spectroscopy}
\author{Alak Ray\altaffilmark{1,2}$^\text{\ddag}$, 
Abraham Loeb\altaffilmark{1}$^\text{\dag}$ 
}
\affil{$^1$Institute of Theory and Computation, Center for
Astrophysics, Harvard University\\ 60 Garden Street, Cambridge, MA 02138}
\affil{$^2$Tata Institute of Fundamental Research, Homi Bhabha Road, Mumbai 400005, India}

\email{$^\text{\ddag}$akr@tifr.res.in}
\email{$^\text{\dag}$aloeb@cfa.harvard.edu}

\begin{abstract}
We propose using radio line spectroscopy to detect molecular absorption
lines (such as OH at 1.6-1.7 GHz)
before and after the total eclipse of black widow (BW) and other short orbital period 
binary pulsars with low mass companions. The companion in such a binary may
be ablated away by energetic particles and high energy radiation produced
by the pulsar wind.
The observations will probe the eclipsing wind being ablated by
the pulsar and constrain the 
nature of the companion and its 
surroundings.
Maser emission
from the interstellar medium stimulated by a pulsar beam 
might also be detected from the
intrabinary medium. The 
short temporal resolution allowed by the millisecond pulsars
can probe this medium with the high angular resolution of the pulsar beam.
\end{abstract}



\section{Introduction}

The first set of planets
orbiting any star other than our Sun were discovered
around a millisecond pulsar ($P_{spin} = 6.2 \;\rm ms$) PSR B1257+20 
at a distance of 600 pc \citep{1992Natur.355..145W}. 
The discovery and follow-up of a similar class of pulsars with low mass companions, 
namely the black widow (hereafter `BW') 
pulsars\footnote{These are defined as a class containing eclipsing binary
millisecond pulsars with ultra-low mass companions}, 
also led to questions of how planets could form or remain around
rapidly rotating pulsars \citep{1993ASPC...36..149P}. It has been argued that
the most plausible scenarios for the formation of these planets
involve a disk of gas around the pulsar \citep{1993ASPC...36..371P}. However
a search for these debris disks around several pulsars such as 
PSR B1257+20 \citep{fos96} from space and ground-based
observatories yielded no detections \citep{2004AJ....128..842L,2014ApJ...793...89W}. 

\begin{figure}
\includegraphics[width=1.0\columnwidth]{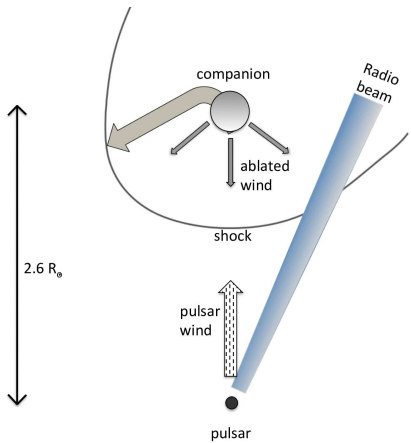}
\caption{Sketch of the geometry for a pulsar 
beam passing through the ablated wind from the super-Jupiter
companion orbiting a black widow pulsar. The fiducial value of
$2.6 R_{\odot}$ for the companion's orbital radius is characteristic of PSR B1957+20
where the inferred radius of the eclipsing region $R_E = 
(f_w c/v_w)^{1/2} R_2 = 0.7 \;\rm R_{\odot}$, where $f_w$ is the
fraction of the incident spin down power on the companion that is
carried off in the form of the wind, $v_w$ is the wind speed, and
$R_2$ is the radius of the ablating companion
\citep{1988Natur.333..832P}. This is a representative case amongst systems with
different orbital periods, pulsar spin down power, and wind parameters. The wind
may be rich in Oxygen if the companion is the remnant of a star that
has generated Carbon or Oxygen during its own evolutionary process.}
\label{fig:eclipse-BWPSR}
\end{figure}

The composition of planets and their atmospheres
around main sequence stars has
emerged as a key area in exoplanet research 
especially after the Kepler and COROT
missions (see \citet{2014prpl.conf..739M} and references therein).  
The discovery of the original black widow pulsar PSR B1957+20 \citep{1988Natur.333..237F,1990ApJ...351..642F},
a millisecond
radio pulsar ablating its companion in a binary system ($P_{orb} = 9.17 \; \rm hr$), 
showed that gas in the eclipsing region is being continually
replenished from the companion's extended atmosphere (see Fig. \ref{fig:eclipse-BWPSR}).
The 1.61 ms pulsar disappears behind the companion in a wide eclipse
for $\sim 10\%$ of its binary orbit. The
companion has a very low mass ($M_c \sim 0.025 \; \rm M_{\odot}$)
and the eclipsing region is substantially larger than the Roche
lobe of the companion.
\citet{1988Natur.334..225K,1988Natur.333..832P} 
suggested that strong gamma-ray irradiation from
the millisecond pulsar drives the wind from the companion
star and gives rise to a bow shock between the wind and
pulsar magnetic field at a distance of roughly ($0.7 R_{\odot}$)
from the companion around which the plasma is opaque to 
radio waves of frequencies $\leq 400 \; \rm MHz$. The pulsar may
be left as an isolated millisecond pulsar as in PSR B1937+21
after few times $10^8 \; \rm yr$. The observability 
of these binary pulsars
in the black widow state implies that the lifetime of this 
transitory phase cannot be much shorter. 
Given the strong pulsar radiation 
and the relativistic electron-positron outflow ablating
its companion to drive a comet-tail like wind, 
it is feasible to search for absorption lines in radio spectra. 
Such lines in absorption spectroscopy
or maser emission in the interstellar medium (ISM) 
has already been detected for several pulsars \citep{2003ApJ...592..953S,2005Sci...309..106W}. 
In this paper, we explore the prospects
of detecting the composition of the gas evaporated from
the very low mass companion of the neutron star, which may in turn
lead to a better understanding of the past evolutionary history of such systems
as well as probe the composition of the companions themselves. This 
in turn could constrain the formation scenarios 
of ultra-low mass companions of pulsars. 
%
%
\section{Black widow and red back pulsars in the galactic field and globular clusters}
The advent of the large area gamma-ray
telescopes like {\it Fermi} soon led to the discovery of
a large population of short period millisecond pulsars
\citep{2014ARA&A..52..211C} that has dramatically increased the number of
black widow like systems similar to PSR B1957+20
\citep{2013IAUS..291..127R}.
The list 
compiled by \cite{2013IAUS..291..127R} was recently supplemented by
PSR J1311-3430 (BW) \citep{2012ApJ...754L..25R} and PSR J2339-0533  (RB) (\citet{2014AAS...22314007R}
and Ray et al (2015, in preparation)). %
Red back pulsars (hereafter `RB') like the black widows
are eclipsing millisecond pulsars with relatively short
orbital periods ($P_B < 1\; \rm d$), but have slightly more
massive companions ($0.1 \; M_{\odot} < m_c < 0.4 \; M_{\odot}$) 
compared to black widow pulsars.
PSR J1023+0038 is the prototype red back pulsar in the Galactic field.
It is a 1.69 ms pulsar in a 4.8 h orbit around a companion star of mass
$0.2 \; M_{\odot}$ whose radio pulsations confirmed it to
be a neutron star \citep{2009Sci...324.1411A}.
Like PSR B1957+20, its eclipse durations are also dependent
on the radio frequency of observation. Its optical studies
combined with Very Long Baseline Interferometry (VLBI) 
observations implied that its companion 
was close to filling its Roche lobe and is non degenerate, 
and suggested that the red backs represented a system
where the neutron star is recently recycled and the accretion\footnote{
This system showed double peaked emission lines in the 
optical spectrum obtained in Sloan Digital Sky Survey in 2001 indicating the presence 
of an accretion disk around the neutron star \citep{2009ApJ...703.2017W}. Whether
red back pulsars evolve to black widow pulsars is however debated
(see \citet{2015ApJ...814...74J,2015ApJ...798...44B}).
} 
is
temporarily halted revealing the radio pulsar which begins to ablate 
the companion 
\citep{2015ApJ...806..148B,2015ApJ...798...44B,2013ApJ...775...27C}. 
PSR J1023+0038 is considered a ``transitional" red back system
since it shows transitions between accretion- and rotation-powered states
and belongs to both classes of red back radio pulsars
and transient low mass X-ray binaries (LMXBs). This pulsar 
provides direct support for the
formation and spinup scenario of millisecond pulsars due
to transfer of angular momentum due to accretion of matter on to the neutron
star from its companion star \citep{1982Natur.300..728A,1982CSci...51.1096R}.
It is the first of this subclass of red backs (the others are:
PSR J1824-2452I, PSR J1227-4853 \citep{2015ApJ...800L..12R} and   
the ``candidate" transitional pulsar
PSR J1723-2837 \citep{2014ApJ...781....6B}.
Black widow and red back pulsars are found also in globular 
clusters\footnote{
See Paolo Freire's compilation \citep{arXiv:1210.3984v3} 
of properties of pulsars in globular clusters: www.naic.edu/$\sim$pfreire/GCpsr.html}.
If the requirement for an eclipse is relaxed, then there are almost as many
millisecond pulsars
with ultra-low mass companion (23) in a globular
cluster as compared to those in the Galactic field (22). 
In the binary containing PSR J1824-2452I in globular cluster M28 there is 
the transition of the opposite kind
which was first
observed as a rotation powered pulsar but then swung to an
accretion powered low luminosity X-ray millisecond pulsar \citep{2013Natur.501..517P}.
High states of $\gamma$-ray emission in the red back PSR J1227-4853
are explained by the Comptonization of disc radiation to GeV energies
by secondary electrons produced in the pulsar slot-gap \citep{2015MNRAS.451L..55B}.
Because of their ever-changing accretion these pulsars may offer
interesting insights towards the formation and evolution of millisecond pulsars
in general and of systems with planetary or ultralow mass companions. 
%
%

For black widows in globular clusters, \citet{2003MNRAS.345..678K} invoked
a two step process, in which cluster turn-off mass stars exchange
into wide binaries containing recycled millisecond pulsars and
remnant helium white dwarf of the donor star, exchanging itself
for the white dwarf and ejecting the latter. Subsequently,
the new companions overflow their Roche lobes because of
encounters and tidal dissipation. The rapidly spinning neutron stars
eject the overflowing gas from the system on a relatively rapid
time-scale at first. The systems enter an observable black widow phase 
at epochs when the evolution is slow, so that the corresponding 
lifetime is long and the probability of finding the system in this phase
is substantial, and the mass loss is small enough
to make the environment transparent to radio waves. The 
incidence of 
known evaporating black widow pulsars among the binary 
millisecond pulsars in globulars and the galactic field are
comparable.
The \citet{2003MNRAS.345..678K} scenario requires high temperature
gas present in the intra and circumbinary region where all molecules
would be dissociated. 
However, \citet{1988ApJ...333L..91W} 
argued the gas temperature to be decreasing adiabatically
with distance from the companion and argued for a much lower temperature
for the gas surrounding the companion of PSR 1957+20 (see below). 
Therefore, detection of a molecular lines from
these systems will 
invalidate the high temperature models.
Unfortunately amongst the pulsars
where L-band flux has been reported, the radio flux density in the 1.6 GHz band
for molecular line detection from almost all black widow and
red back pulsars in globular clusters 
are too low to be observable with the current generation of radio
telescopes, except possibly PSR J1807-2459A, but several may become
detectable with the greater sensitivity of the Square Kilometer Array (SKA). 

\cite{2011Sci...333.1717B} have discovered a millisecond pulsar
PSR J1719-14 in a very close binary system ($P_{orb} = 2.2 \; \rm hr$)
whose companion has a mass near that of Jupiter, but its minimum
density suggests that it may be an ultra-low mass carbon
white dwarf. This system does not show any evidence of
solid body eclipses or excess dispersive delays perhaps due
to an unfavorable angle of inclination of its orbit relative to our line
of sight. Nevertheless, its discovery points to the existence of
systems which may have once been an ultra compact low mass 
x-ray binary where the companion has narrowly escaped
complete evaporation. Similar systems 
with more favorable inclination
angles may be discovered in future surveys with wider coverage
and computational analysis power expected in the SKA\footnote{
See 
\citet{2015aska.confE..40K,2015aska.confE..39T}
in The SKA Key Science Workshop Documentation at: 
http://astronomers.skatelescope.org/meetings-2/2015-science-meeting/}
era and will be of
interest 
for the reasons we discuss next.

\begin{deluxetable*}{l c c c c c c c c c c}
\tablecolumns{11} 
\tablewidth{0pc} 
\tablecaption{Black widow and other pulsar targets for OH line spectroscopy\tablenotemark{A}} 
\tablehead{ 
Pulsar & $l, b$ & $\rm P_{spin}$ & W50 & $\rm P_B$ & Min. $\rm m_c$ & Age\tablenotemark{B} & $\rm \dot E$ & D & $S_{1400}$& Ref. \\
       &  deg   & ms & ms & hrs   & $\rm M_{Jup}$ & $10^9$ yr & $\rm L_{\odot}$ & kpc & mJy &   }
\startdata
J2051-0827 &  $39.2, -30.4$& 4.51 & 0.34 & 2.4 & 28.3 & 5.61 & 1.4 & 1.28 & 2.8 & $a$, $b$, $c$\\ 
J2241-5236 &  $237.5, -54.9$& 2.19 & 0.07 & 3.4 & 12.6 & 5.22 & 6.5 & 0.68 & 4.1 & $d$\\ 
J0751+1807 &  $202.7, -21.1$& 3.48 & 0.70 & 6.3 & 134\tablenotemark{D}  & 7.08 & 1.9 & 0.40 & 3.2 & $d$,$h$,$i$\\ 
J1012+5307 &  $160.3, +50.8$& 5.25 & 0.69 & 14.4& 112\tablenotemark{D}  & 4.86 & 1.2 & 0.70 & 3.0 & $d$,$j$,$k$\\
J1807-2459A\tablenotemark{C}& $5.8, -2.2$& 3.06 & 0.33 & 1.7 & 9.4  & --   & --  & 2.79 & 1.1 & $e$,$f$,$g$\\
B1259-63   &  $304.2, -1.0$& 47   & 23   & 1236d & $10 M_{\odot}$ & $3\times 10^5 \rm yr$ & 207 & 2.3 & 1.7 & $l$, $f$, $m$\\
B1957+20   & $59.2, -4.7$ & 1.61 & 0.035 & 9.1 & 22 & 2.2 & 40 & 1.53 & 0.4 & $n$, $c$ \\
J1227-4853\tablenotemark{F} & $298.9, +13.8$ & 1.69 & ----- & 6.8&221\tablenotemark{D} & 2.4 & 24 & 2.00 & ---\tablenotemark{F} & $r$ \\
J1723-2837 & $357.6, +4.26$ & 1.86 & 0.20 & 14.8 & 247\tablenotemark{D} & 3.9 & 12 & 1.00 & 1.1 & $s$ \\
\label{obstab}
\enddata
\tablenotetext{A}{Pulsars with moderate {\it mean} flux density S1400; from ATNF Pulsar Database \citep{2005AJ....129.1993M}:
$\rm www.atnf.csiro.au/people/pulsar/psrcat/$; PSR B1957+20 is
included
for comparing parameters}
\tablenotetext{B}{$\rm Age = P_{spin}/2\dot P$; $\rm \dot P$ uncorrected for Shklovsky effect} 
\tablenotetext{C}{in Globular Cluster NGC6544; pulsar
spindown age and luminosity are uncertain
due to acceleration effects}
\tablenotetext{D}{these companions correspond to red back pulsars}
\tablenotetext{F}{estimated flux in 1.4 GHz band is $1.6 \rm mJy$ using
the reported 607 MHz flux and spectral index -1.7 from \citep{2015ApJ...800L..12R}}
\tablenotetext{a}{Stappers et al (1996)}
\tablenotetext{b}{Doroshenko et al (2001)}
\tablenotetext{c}{Kramer et al (1998)}
\tablenotetext{d}{Keith et al (2011)}
\tablenotetext{e}{Ransom et al (2001)}
\tablenotetext{f}{Hobbs et al (2004)}
\tablenotetext{g}{$\rm Lynch$ et al (2012)}
\tablenotetext{h}{$\rm Lundgren$ et al (1995)}
\tablenotetext{i}{Nice et al (2005)}
\tablenotetext{j}{Nicastro et al (1995)}
\tablenotetext{k}{$\rm Lazaridis$ et al (2009)}
\tablenotetext{l}{$\rm Johnston$ et al (1992)}
\tablenotetext{m}{$\rm Melatos$ et al (1995)}
\tablenotetext{n}{$\rm Fruchter$ et al (1988)}
\tablenotetext{r}{Roy et al (2015)}
\tablenotetext{s}{Crawford et al (2013)}
\end{deluxetable*}

\section{Planets around pulsars, their evaporation and intrabinary gas}

Since the discovery of the first extra-solar planet
around PSR B1257+12 \citep{1992Natur.355..145W} there has been only
one other radio astronomical discovery of a super-Jupiter planet in a
triplet system within a globular cluster namely PSR B1620-26 
\citep{1993ApJ...412L..33T, 2003ApJ...597L..45R,
2003Sci...301..193S}.
Despite sensitive monitoring of more than
151 young ($\tau < 2 \; \rm Myr$), luminous pulsars
for periodic variation in pulse arrival time due
to possible planetary companions, \cite{2015ApJ...809L..11K} failed to detect any
further planetary companions around pulsars. 
However these pulsars are not only young and slow rotators, but their
pulse timing properties are usually far less 
accurate than that for old millisecond pulsars hosted by black widow systems.
It is therefore possible that in addition to the causes mentioned by 
\citet{2015ApJ...809L..11K} the search could have been less sensitive for
planets around normal pulsars as opposed to those around millisecond pulsars like
PSR B1257+12.
\citet{1993ASPC...36..149P,1993ASPC...36..371P}
classified the various
planet formation models for planets around pulsars (especially
in the context of PSR B1257+12) into
two groups: i) the planets formed around an ordinary star like
in our solar system and later on this star exploded and created
a spinning neutron star, yet the planets survived, or the planetary
system was captured by a neutron star in a direct collision with the
solar-type main sequence central star; and ii) the planets formed
soon after the pulsar was born in a supernova explosion. 
\citet{1993ASPC...36..149P} suggested a third set of models
in which planet formation constitutes the final stage in the evolution
of some millisecond pulsars, e.g. the circumbinary disk models,
where a binary companion of a millisecond pulsar is being evaporated
\citep{1993ApJ...415..779B}, or in which the evaporation is taking place in a
low mass X-ray binary phase \citep{1992Natur.356..320T}. In contrast to the
simple evaporation models, some of the material may not escape
from the system and may form a circumbinary disk, from which planet formation
can take place. 

The environment of the planets around pulsars is characterized by high 
irradiation of
both 
photons and particles 
from the millisecond pulsar.
Chemistry in this gaseous medium 
may bear similarities to dense photon dominated regions \citep{1995ApJS...99..565S},
where more energetic photons than far-ultraviolet (FUV) radiation as well as electron and
baryonic particle flux may be substantial. As shown by \citet{1995ApJS...99..565S}, 
various gas-phase photochemical processes may lead to the production of atomic and molecular
species in dense photon dominated regions. 
The physical and chemical
properties of photon dominated regions in a dense molecular cloud depend upon the gas density
and pressure, intensity of high energy radiation,
gas phase elemental abundances and the presence of dust grains. 
%
\citet{1995ApJS...99..565S}
showed that  $OH$, $H_2O$
and their ionic forms or their precursor molecule $H_3O^+$ may form in both photon
mediated as well as electron mediated processes in hot $HI$ zone and near the $H/H_2$
transition layer (see section 3.1.1 of their paper).
 They employed models involving
a static, plane parallel, semi-infinite cloud exposed on one side to an
isotropic radiation
field with a constant hydrogen particle density $n_T = n_H + n_{H_2} = 10^6 \; \rm cm^{-3}$ 
throughout the cloud and an incident FUV field with an intensity $2 \times 10^5$ times
the average interstellar FUV field estimated by \citet{1978ApJS...36..595D}. They found
that the OH abundance, a crucial intermediary in the chemistry of the hot gas that leads to
the production of many molecules and molecular ions,
reaches a maximum at a visual extinction $A_V = 0.6$ where the gas temperature is  $800\; \rm K$.
In fact their density ratio with respect to $H_2O$, $OH/H_2O$ dominates even at higher
visual extinctions up to $A_V =3$ in the clouds. With the assistance of reactions on grain
surfaces, hydrogen molecules form by the association of hydrogen atoms 
and the neutral hydrogen HI gas is hot with temperatures exceeding $10^3\; \rm K$
at a depth $A_V = 0.7$ from the cloud surface. 
This is because the gas heating by grain photoelectric emission
and collisional de-excitation of energetic radiation-pumped $H_2$ becomes efficient
while emission cooling is quenched \citep{1995ApJS...99..565S,1990ApJ...365..620B}. 

The environment of a pulsar which resulted from a supernova explosion is likely to be
particularly rich in metals. Grain formation can take place in metal rich gaseous surroundings
of the pulsar and its planet.
Much of the chemistry in planet forming regions
around normal stars has been shown recently to be driven by gas-grain chemistry. The smaller grains
of size between 
$0.001 \rm \mu m$  and $0.1 \rm \mu m$ 
which provide most of the surface area for 
chemistry are critical for absorption and scattering of UV radiation \citep{2014FaDi..168....9V}.
Interstellar grains are agents through which surface molecules
participate in promoting the reaction rather than having an active role as catalysts.
Primarily, they provide a reservoir where atoms and molecules can be stored and brought closer
together for much longer periods than probable in the gas phase and can enable reactions
that are too slow with substantial activation barriers. Additionally,
by acting as a third body that absorbs the
binding energy of the newly formed molecule, they stabilize it from dissociating quickly.
As in canonical planetary systems, long lived pulsars with ultra-low mass 
companions may have disks where grains form and assist molecule formation.
\medskip

\section{Radio absorption spectroscopy of ablated gas from the planet or companion object}
\smallskip
The time dependent
nature of the pulsar emission offers 
an advantage: the narrow beam of pulsar
radiation makes a pencil sharp probe of the intervening medium especially
if  it gives rise to an absorption dip in the pulsar continuum. The
pulsar ``on" spectrum represents the signal of the pulsar alone as modified
by absorption and in rare circumstances by stimulated emission \citep{2005Sci...309..106W} 
by the intervening medium in a narrow angle.
In contrast, the pulsar ``off" spectrum in the intervening time between the
pulses has both line emission and absorption occurring
within the 
wider telescope beam. 
Given the
flux density of most pulsars fall off rapidly with observing radio frequency
it is natural to target pulsars for atomic  or molecular line transitions
in the 
L- (1.4 GHz) band and possibly in the X- (8-12 GHz) band.
\cite{2003ApJ...592..953S} detected molecular OH absorption against PSR B1849+00
at 1.665 GHz and at 1.667 MHz. 
In addition, 
there exist other
hyperfine transition lines of $^3He^+$ ($^2S_{1/2}, F = 0-1$)
at 8.665 GHz 
and molecular lines Ortho-Formaldehyde ($H_2CO$) at 4.83 GHz
and of Methanol ($CH_3OH$) at 6.67 GHz with relatively large
Einstein A coefficients \citep{1996tra..book.....R}. Although the Helium isotope
$^3He$ has an abundance of $1.38\times 10^{-4} \; \%$ ($^4He$ abundance is
likely to be high in an evolved remnant of companion of the pulsar), since the
pulsar spectra decline steeply with frequency, it might be challenging at present
to detect these atoms or molecules in absorption or stimulated emission against pulsar radiation. 

\smallskip
\subsection{Molecular gas: interstellar or in situ binary?}

\citet{2003ApJ...592..953S} and \citet{2005Sci...309..106W} detected OH absorption in PSR B1849+00
and PSR B1641-45 using Arecibo and Parkes radio telescopes respectively
out of a total of 25 pulsars. In addition \citet{2008ApJ...677..373M} 
detected OH absorption for PSR B1718-35 by using the Green Bank
telescope (GBT) out of a sample of 16 pulsars. While 
PSR B1641-45 has a relatively large mean flux density
of $S_{1400} = 310 \; \rm mJy$, the other two have 
11 mJy (PSR B1718-35) and 2.2 mJy (PSR B1849+00) respectively.
Thus, typically only about 10\% of the searched for targets show evidence of
OH absorption (or stimulated emission) in pulsar lines of sights. The 
detections have been made only for low-galactic latitude pulsars
(see galactic $(l,b)$ in Table \ref{OH-absorption-detections}).
However almost all
target pulsars listed in Table \ref{obstab} turn out to be high galactic latitude pulsars. 
Since OH is generally strongly confined to the galactic plane, any detection of OH absorption
(or stimulated emission) in these systems will be related to the binary system with high probability rather than
be interstellar. 
The intrabinary molecular gas signature can be distinguished from circumbinary molecular gas
by the orbital phase modulation (i.e. when the pulsar is near inferior conjunction).
Binning the recorded data with
respect to companion orbital phase would 
allow the exclusion of
low signal to noise (for molecular gas) in 
high impact parameter data (when companion orbital position is far away from 
inferior conjunction).
In case of very high signal to noise of atomic/molecular gas 
in the background of a bright pulsar it may even be possible to study the
variation of absorption and emission with the impact parameter which can then
lead to information about the radial structure of the ablated wind from the companion.
As only times close to the total eclipse would have substantial OH absorption in the trailing
wind, only the fraction of the orbit near the eclipse ingress and egress have to be
added from multiple orbital cycles.

%

\subsection{OH molecular line absorption towards pulsars}
\smallskip

In the context of the
black widow PSR B1957+20 one can 
expect
sufficiently large column depths of OH molecules near the
eclipsed region, i.e. in orbital phases close to the inferior conjunction
to show moderate
line absorption with reasonable equivalent width (EW).
This is because the gas can cool adiabatically with distance
from the ablating companion star (see below) to enable molecule formation.
Unfortunately the mean flux density ($\rm S1400 = 0.4 \; \rm mJy$) for this pulsar
is too low for an easy detection by
existing telescopes.
%
%
Table \ref{obstab} lists the characteristics of pulsars that may be
suitable for detection of molecular lines through radio spectroscopy.
Note that we include here
PSR B1957+20 for comparison -- because of its very low mean flux density
at 1400 MHz 
it is not meant to be a target for OH detection with the current generation
of telescopes. 
On the other hand PSR B1259-63 has a massive Be star companion 
which does not belong to the black widow or red back classes but whose 
high $\Delta DM$ post- and pre-eclipse phases last for tens of days
during which an OH line spectroscopy can be carried out.
Table \ref{OH-absorption-detections} lists pulsars from which
OH line absorption has been observed in the 
ISM
during the pulse ``on" phase. For PSR J1641-45, not only line
absorption at 1612 MHz, 1665 MHz and 1667 MHz, but even stimulated
emission at 1720 MHz driven by the pulsar during its pulse ``on" phase
has been detected by \cite{2005Sci...309..106W}.
This is the only pulsar, an extremely bright one, from which the stimulated
emission was discovered, not among the other fainter ones in Table \ref{OH-absorption-detections}.
Therefore, the use of stimulated emission to probe the conditions of the companion's
wind in a binary pulsar may be possible if such an extremely bright pulsar were to be
discovered in a binary system.
The examples in Table \ref{OH-absorption-detections} give column depths
of OH line absorption
already detected against isolated and bright pulsars
with existing telescopes in reasonable exposure times. They are ``ordinary" pulsars
with duty cycles ($ = \rm W50/P_{\rm spin}$) ranging from $1.8\%$ to $10.8\%$,
moderate radio flux densities in the 1400 MHz band close to where
the OH line absorption at 1612-1720 MHz band is expected and
are young pulsars with spindown power comparable to the
solar luminosity. The targets listed in Table \ref{obstab} are {\it binary, millisecond} 
pulsars with high spindown power and
moderate radio flux densities at 1400 MHz,
with ultra-low mass (or even comparable to Jupiter mass) companions
in close orbits ($P_{orb} \sim 2- 14 \; \rm hr$). Note their duty cycles
are comparable to those in Table \ref{OH-absorption-detections} even though
they are much more rapidly spinning and their moderate
mean flux densities at 1400 (1600) MHz and their
small orbital dimensions make them good targets
for HI/OH absorption studies against their pulsed flux. Black widows
among the entries in Table \ref{obstab} (the first two) have 
ultralight companions that are partially degenerate and stripped due to ablation by the pulsar.
X-ray studies of
black widow pulsars show that their non-thermal X-ray emission
are orbitally modulated which has been related to the
intrabinary shocks close to the companion \citep{1988Natur.333..832P,2013IAUS..291..127R}.  

\begin{deluxetable*}{l c c c c c c c c c c c}
\tablecaption{Pulsars with OH absorption detected against pulsed emission} 
\tablehead{
Pulsar & l, b &$\rm P_{spin}$& W50 & Age$^A$ & $\rm \dot E$    & D   &$S_{1400}$&$N_{OH}/T_{ex}$&Telescope&Exp.&Ref.\\
       & degree&     ms     & ms  & $10^5$ yr&$\rm L_{\odot}$&kpc& mJy &$10^{14}\rm cm^{-2}K^{-1}$& &hrs& }
B1641-45 &$339.2, -0.2$& 455& 8.2 & 3.59 & 2.2 & 4.5 & 310  & 0.28\tablenotemark{B} &Parkes& 5& $1$\\ 
B1718-35 &$351.7, +0.7$& 280& 26  & 1.76 &11.8 & 4.6 & 11.0 &0.1-0.67 &GBT&$< 10$ & $2$\\ 
B1849+00 &$33.5, +0.02$&2180& 235 & 3.56 &0.1  & 8.0 &  2.2 &2.7 &Arecibo& 3.4& $3$\\ 
\label{OH-absorption-detections}
\enddata
\tablenotetext{A}{$\rm Age = P_{spin}/2\dot P$. Note the different units of age 
used in Table 1.}
\tablenotetext{B}{Calculated from $\tau = 0.03$ at 1667 MHz with FWHM $\Delta v = 2 \;\rm km s^{-1}$ reported in Ref. 1.} 
\tablenotetext{1}{Weisberg et al (2005)}
\tablenotetext{2}{Minter (2008)}
\tablenotetext{3}{Stanimirovic et al (2003)}
\end{deluxetable*}

The eclipse duration in PSR B1957+20 is a function
of observing radio frequency. \citet{1989ApJ...342..934R} and
\citet{1988ApJ...333L..91W} argued that the radio radiation is being absorbed
with a frequency dependent cross section rather than
refraction and reflection from ionized plasma wind \citep{1988Natur.333..832P}. 
Radio line absorption in the OH 1.6 GHz band against pulsed flux
can be easily associated with interstellar vs in-situ (planetary or
intra-binary) origins by the independent evidence of binarity and
a correlated change of OH absorption with orbital phase near
eclipse ingress and egress. 

The observed line intensity profile of OH absorption 
can be written in terms of the 
OH column density
$N_{OH}$ as \citep{2006ApJ...652.1288H}):

\begin{align}
&I_{\nu} = (1-f) I_{0\nu} \nonumber \\ 
&        + f\{I_{0\nu}\; \rm exp(-\tau_{\nu}) + B_{\nu}(T_{ex})[1 - \rm exp(-\tau_{\nu})]\} \nonumber 
\end{align}
where $I_{0\nu}$ is the background continuum intensity, represented
by the pulsed radio flux during the pulsar ``on" phase, $B_{\nu}(T_{ex})$
is the Planck function, $T_{ex}$ is the excitation temperature
of the four transitions that represent the OH molecular lines in the L-band,
$\tau_{\nu}$ is the optical depth at frequency $\nu$ and $f$ is a telescope
beam filling factor (which we take to be unity as only the pulsar ``on"
phase is compared with the flux in between the pulses, i.e. ``off-pulse").
This can be rewritten in terms of the brightness temperature, assuming 
that $T_{ex} \gg \rm h \nu_{OH}/k_B = 0.08 \; \rm K$ where the Rayleigh-Jeans
approximation holds:

$$ T(v) - T_0 = f (T_{ex} -T_0)\{1 - \rm exp\;[-\tau(v)]\}.$$ 
Here $\tau(v)$ can be written in terms of the optical depth at the line center
$\tau_0$
with a Gaussian profile:
$$ \tau(v) = \tau_0 \; \rm \exp[ -(v-v_0)^2/(2\sigma^2)]$$
where the line width $\sigma$ is related to the velocity dispersion $\Delta v$ through 
$\sigma = \Delta v/ (8 \rm ln 2)^{1/2}$.
The optical depth at the line center depends upon the OH molecular column
density $N_{OH}$, corresponding Einstein A-coefficient, degeneracies of
the upper and lower states and the occupation fraction of the initial state, 
and the excitation temperature $T_{ex}$ and can be collectively written as:
\begin{equation}
\tau_0 = a \frac{N_{OH} (10^{15} \; \rm cm^{-2})} {T_{ex}(\rm K) \Delta v (\rm km s^{-1})}
\end{equation}
with the constant $a =$ 0.454, 2.345, 4.266 and 0.485 for
the 1612, 1665, 1667 and 1720 MHz lines respectively (see \cite{2006ApJ...652.1288H}). 
For the strongest ``main" line at 1667 MHz, this leads to the
column depth:
\begin{equation}
N_{OH} = 2.3 \times 10^{14} \rm cm^{-2} \; (T_{ex}/\rm K) \int \tau(v) dv/(\rm km s^{-1})
\end{equation}
As an example, \citet{2003ApJ...592..953S} obtained for {\it interstellar} OH absorption
against PSR B1849+00, $\tau_{max} = 0.4 \pm 0.1$
at 1665 MHz and $\tau_{max} = 0.9 \pm 0.1$ at 1667 MHz with corresponding
line-width FWHM of $1.5 \pm 0.4\; \rm km \; s^{-1}$ and $1.1 \pm 0.2\; \rm km \; s^{-1}$. 
The corresponding $N_{OH}/T_{ex}$
are tabulated in Table \ref{OH-absorption-detections}.

\smallskip
\subsubsection{Eclipsing gas temperature}

What is the temperature of the eclipsing gas at
a pulsar line of sight impact parameter $b$? Here $b= \pi a \Delta \phi$, where
a is the radius of the orbit of the planet or ultra-low mass stellar companion
and $\Delta \phi \ll 1 $ is the orbital phase of the eclipse.
In the context of the original black widow pulsar
PSR B1957+20, \citet{1988ApJ...333L..91W} found the gas temperature to
be $\sim 260\; \rm K$ from the dispersion
measure fluctuations accompanying emersion from the eclipse,
assuming that the eclipse is caused by free-free optical
depth. Temperature is expected to decline adiabatically as $T \propto r^{-4/3}$
with distance $r$ from the 
companion.
Surface temperatures of the companions (the optical counterparts) of a number of black widow and
red back pulsars (detected in $\gamma$-rays with Fermi Gamma Ray Telescope)
were determined by UV and optical observations (Gemini, NTT and Swift-UVOT)
by \citet{2013ApJ...769..108B}. They measured both ``dayside" and ``nightside" temperatures (i.e.
of the hemispheres
of the companion facing the pulsar or on its opposite side). While ``nightside" temperatures can be
of the order of $\sim 2500$ K the ``dayside" temperatures range between $4200-8000$ K. 
\citet{2013ApJ...765..158K} in their Fig 6 compare 
surface temperatures and gravity of the companions of several black widow and red back pulsars,
e.g. in PSR J0751+1807 (with $m_c \sim 0.14 M_{\odot}$)
temperature could be as low as
$3500 \; \rm K$ \citep{2006A&A...450..295B}.
HST observations of the companion of PSR J2051-0827 have yielded
temperatures $\sim 2500 \; \rm K$ on its unirradiated side in certain orbital phases
\citep{2001ApJ...548L.183S}. 
If the companions are very low mass proto-He white dwarf companions of the millisecond pulsars
that may be out of Roche lobe contact for over several Gyr, their radius
$R_c$ contracts to $\leq 0.07 R_{\odot}$ (see Fig 3 of \citet{2014A&A...571L...3I}; on
the other hand, for
a core derived from a main sequence star with cosmic abundance, the companion radius
$R_c \sim 0.1 R_{\odot}$ as in the case of PSR B1957+20). 
With the radius and surface temperature of the He white dwarf companion, gas
on the nightside cooling adiabatically with $T \propto r^{-4/3}$
would reach $\sim 100 \;\rm K$ within a distance $16 R_{c} \sim 1 \; R_{\odot}$
from the companion, which is 
the typical 
dimension of the eclipsing region 
in a black widow pulsar \citep{1988Natur.333..832P}.

\smallskip
\subsubsection{OH column depths}

To estimate the column depth of $N_{OH}$
from beyond the eclipsing region 
we may adopt $T_{ex} \sim 100 \; \rm K$ and a 
line-width
of FWHM $\sim 3 \; \rm km\;s^{-1}$ (see section 5.1). Since the ablated wind
from the companion is primarily in the tangential direction,
or may reside for long periods in a Keplerian disk centered on the
neutron star,
the effective velocity width of the radiatively connected region
in the line of sight along the pulsar beam may not exceed much 
the thermal speed. Therefore, a
velocity resolution of 
$1.0 \; \rm km \; s^{-1}$
would sufficiently resolve the absorption profile in several bins.
Adopting
$\tau_0 = 0.3$ at 1667 MHz (similar to that
observed in {\it interstellar}
OH absorption against PSR B1718-35 with a comparable spectral resolution
as in \citet{2008ApJ...677..373M}),
one requires an intra-system column depth of
$N_{OH} = 2.28 \times  10^{16} \times 0.9 \sim 10^{16} \rm \; cm^{-2}$.

We 
argue that
this 
column density 
is comparable to what is 
derived from the 
mass loss rate from the pulsar companion in a wind outflow:
$$ \dot M_w = f_{\Omega} 4 \pi \rho r^2 v_w .$$
Here $f_{\Omega}$ is the fraction of the total solid angle covered
by 
the
wind ablated from the companion 
$\rho$ is the mass density in the wind at a distance $r$ from the
pulsar and $v_w$ is the speed of the wind being blown out by the
pulsar energetic outflow in photons and particles.
The column density $N_{OH}$ that radio radiation from the pulsar
encounters 
is:
$$N_{OH} \sim (X_{OH}/m_{OH}) \rho \times s .$$ 
Here, $X_{OH}$ is the OH fraction\footnote{The Solar Oxygen nucleon fraction 
is $X_O = 5.5 \times 10^{-3}$
using the scaling $n_O/ n_H = 4.89 \times 10^{-4}$ from 
$\log \; \epsilon_O = 8.69$ \citep{2009ARA&A..47..481A} with the definition:
$\log \; (n_X/n_H) +12 = \log \; \epsilon_X$.} in the wind (fraction
of the nucleons locked up in OH), whose typical value is
$\sim 10^{-3}$, 
$n_X$'s are the number {\it densities} of the
element X, and $s$ is the absorption lengthscale encountered by the pulsar beam in the 
intrabinary space
typically of dimensions comparable to that of the eclipsing region, 
and $m_{OH}$ is the mass of an OH molecule.
For the case of PSR B1957+20,
\citet{1988Natur.333..832P} has argued that the eclipsing region is $\sim 0.7 R_{\odot}$,
and so we adopt in our calculations $s \sim r$.
The expression above can be further reduced in terms of parameters of
the pulsar system under consideration by the equating the timescale
of complete ablation of the planetary mass companion to the spin down timescale
for the pulsar:
\begin{equation}\label{eqn:NvsMdot}
N_{OH} = \frac{\dot M_w X_{OH} s} {f_{\Omega} 4 \pi \rho r^2 v_w m_{OH}}.
\end{equation}
In terms of the parameters of 
PSR B1957+20 with an assumed evaporation timescale\footnote{The timescale
of ablative mass loss from the system can be indirectly calculated from the 
orbital period evolution \citep{1988MNRAS.235P..33C,1990ApJ...351..642F}.
However,
The orbital period of PSR B1957+20 is undergoing small, apparently random
variations on 5 yr timescale \citep{2000ASPC..202...67N} which {\it may be} related to
the quadrupolar deformation of the magnetically active secondary with a 
sizeable convective envelope. The secondary may undergo a wind driven mass loss, 
which 
is powered by tidal dissipation of
energy and a torque on the companion \citep{1994ApJ...436..312A}.
\citet{2011MNRAS.414.3134L} argue that for PSR J2051-0827
gravitational quadrupole and classical spin-orbit coupling
can together account for its observed
orbital variations if the companion is under-filling its Roche-lobe
by only a moderate factor, e.g. with a radius $R_c =0.14 R_{\odot}$.
The orbital
period change of millisecond pulsar PSR J1744-24A 
implies a
timescale of $|P_{\rm orb}/\dot P_{\rm orb}|_{\rm obs} = 200\; \rm Myr$
and the residual pulsar timing noise in these binary systems is interpreted
as due to mass flow in the system \citep{2000ASPC..202...67N}.} 
yields a column density,
%
\begin{align}
& N_{OH} = 4.8\times 10^{16} \rm cm^{-2} \Big(\frac{m_c}{0.02 \rm M_{\odot}}\Big)\Big(\frac{X_{OH}}{10^{-3}}\Big)\Big(\frac{s}{r}\Big) \nonumber \\
& \times \Big[\Big(\frac{\tau_{evap}}{2\times 10^8 \rm yr}\Big)\Big(\frac{f_{\Omega}}{0.25}\Big)\Big(\frac{r}{2.6 \rm R_{\odot}}\Big)\Big(\frac{v_w}{200 \rm km s^{-1}}\Big)\Big]^{-1} 
\end{align}
\smallskip

This column density is indeed
comparable to the values measured from the observed interstellar OH absorption towards
pulsars (see Table \ref{OH-absorption-detections}). 
\smallskip

To assess the column depth of $N_{OH}$ in other binary pulsar systems listed in Table \ref{obstab},
we note that variables in Eq. (\ref{eqn:NvsMdot}) can be scaled to observable
or listed parameters in Table \ref{obstab}. 
As before, we assume that 
$(s/r) \sim \rm constant$. Furthermore
\citet{1988Natur.334..225K} 
derive $\dot M_w$  as well
as the evaporative wind velocity $v_w$ driven by total 
$\gamma$-ray radiation and TeV $e^{\pm}$ pairs radiated by the pulsar and intercepted
by the secondary 
when the binary pulsar enters
the present phase of radio pulsar driven evaporation of the secondary 
in PSR B1957+20. Using these expressions, we estimate:
$N_{OH} \propto L_{\gamma}^2  r^{-5} h R_c^2$, where $L_{\gamma}$ is the Mev $\gamma$-ray 
luminosity, $R_c$ is the companion radius intercepting a fraction of 
that luminosity at an orbital distance $r$, and $h$ is the scale height of the companion's
(inflated) atmosphere which may extend to a significant fraction of the Roche lobe. 
Since the engine 
of both the gamma-ray luminosity as well as the TeV $e^{\pm}$ luminosity is the spin down power
of the neutron star, we can scale the column density of molecular OH in terms of the
pulsar parameters as:
$N_{OH} \propto \dot E^2/ P_{orb}^{10/3}$.
For parameters listed in Table \ref{obstab}, the highest $N_{OH}$ is
predicted for the original black widow pulsar PSR B1957+20, but the $N_{OH}$ 
for the first 3 entries of Table \ref{obstab} are within an order of magnitude of that predicted for
PSR B1957+20. 

The red back "candidate transitional" pulsar
PSR J1723-2837 has a mean flux density (at 1400 MHz) of 1.1 mJy as
listed in ATNF catalog. However with an orbital period 14.76 hrs
the column density of $N_{OH}$ could be low.
This pulsar may be a good candidate in the upcoming SKA era for attempts to detect
OH/HI lines, but is likely to be challenging with present telescopes if its 
flux remains steady. The transitional millisecond pulsar PSR J1227-4853
\citep{2015ApJ...800L..12R} with shorter orbital period and
significant $\dot E$ has been
found to eclipse at 1420 GHz (Parkes) and
at 607 MHz (GMRT). Its continuum flux density at 607 MHz
is 6.6 mJy; assuming a
spectral index of -1.7 used by \citet{2015ApJ...800L..12R}, the estimated flux at 1420 MHz would be
1.6 mJy, which may also pose a challenge 
for atomic and molecular
line detections even if its column depth of molecular OH could be significant.
\smallskip
\subsection{Atomic Hydrogen HI in the binary system?}
\smallskip
%
If both molecular OH and neutral atomic 
hydrogen are 
present in the gas ablated from the ultra-low mass companion of the energetic pulsar,
one can also estimate the possible column depths.
If the 
$X_{OH}$ fraction is slightly subsolar, e.g. $\sim 10^{-3}$ as 
we have assumed above,
the column depth of HI in the intrabinary
gas, assuming much of the gas is made of hydrogen
and hydrogen is locked up mostly in atomic form (HI) once the gas cools down to
temperature $\sim 100 \; \rm K$, and 
scaling the $N_{OH}$ column
depth in Eq. (\ref{eqn:NvsMdot}) yields $N_{HI} \sim  (5-10) \times 10^{19} \; \rm cm^{-2}$.
Given the relation between column depth and the EW for HI absorption \citep{1988gera.book...95K}:
$ N_{HI} = 1.8 \times 10^{18} \langle 1/T_s\rangle^{-1} EW$
where $\langle 1/T_s \rangle^{-1}$ is harmonically weighted spin temperature along the path, 
whose value may be 50-100 K,
one finds $EW \sim 2 \; \rm km \; s^{-1}$. 
One can detect such EW's in the {\it pulsar spectrum} easily.
Note that variations in column densities of $\sim 10^{18-19} \; \rm cm^{-2}$
have been detected on occasions towards strong pulsars like PSR B1929+10
\citep{2007ASPC..365...28W,2010ApJ...720..415S}.
This gives an indication of the scale of measurable HI column depths in ISM and
the estimated $N_{HI} \sim  (5-10) \times 10^{19} \; \rm cm^{-2}$ derived from Eq. (\ref{eqn:NvsMdot})
can therefore be detected. 

\smallskip
\section{Observational requirements} 

The strategy for the detection of molecular line absorption (or emission) 
in the intervening medium probed by the
narrow beam of pulsar radiation involves two primary considerations. 
First, the pulsed flux of radiation should be high 
enough for easy detection in narrow radio frequency bands that are fine enough 
to sample the absorption line {\it width}
sufficiently well. 
The linewidth is determined by a combination of thermal broadening and
microturbulence in the absorbing gas, -- the latter is due to  many small cells of gas moving
in random directions with a Maxwellian distribution of speeds,
leading to a Gaussian line profile (which is however
independent of the mass of the absorbing atom or molecule, in contrast to the
thermal width which varies
$\propto m^{-1/2}$). 
Moreover due to binary companion's motion and the speed of the ablated
tail wind, the centroid of the spectral line may move over a relatively wider range, 
compared to thermal line width.
Even though one would like to determine the line profile,
the narrow line width is a challenge
since resolving the line requires high spectral resolution
and given a pulsar flux density, the required signal to noise for pulse detection
scales with frequency bandwidth
as: $\propto (\Delta f)^{1/2}$. However, even when a line profile cannot be measured
with great precision due to spectral resolution issues, 
equivalent width\footnote{
The equivalent width, $W_{\nu}$, or $W_{\lambda}$,
that measures the strength of the absorption line is the width of
the adjacent continuum that has the same area (in the plot of
the radiance per unit frequency or wavelength vs frequency or wavelength) as taken up by the absorption
line.
$W(\lambda_0)
= \int_{-\infty}^{\infty}[1 - \exp(-\tau (\lambda - \lambda_0))] d\lambda$,
and 
$W_{\nu_0} = (c/\lambda_0^2) W_{\lambda_0} = \int_{-\infty}^{\infty}[1 - \exp(-\tau (\nu - \nu_0))] d\nu$.
When $\tau_{\nu}$ is small across the line, as in the optically thin case,
the equivalent width is linearly proportional to the column depth of the absorbing atoms or molecules
along the line of sight \citep{1978ppim.book.....S}.
} 
and curve of growth analysis \citep{1978ppim.book.....S} have been used
to deduce gas temperature, column density and abundance of elements. 
Second, the sensitivity of the spectral measurement should be high enough to 
adequately sample the flux inside the absorption
{\it depth} for the expected column of the atoms or molecules along the line of sight. 
That is, it should yield the 
equivalent width
of the absorption line with sufficient accuracy.
The sensitivity to 
measuring small equivalent widths of absorption lines improves directly with
resolving power, provided adequate signal to noise can be achieved in the
continuum neighboring the lines \citep{2005hris.conf....3B}.
%

\subsection{Equivalent width of the absorption line}
\smallskip
Measurement of absorption spectra of interstellar
HI and OH against the pulsed radiation has been described in 
\citet{1995ApJ...441..756K,2003ApJ...592..953S,2005Sci...309..106W} (see especially the supporting
on-line material of the last reference). Briefly, 
a correlation spectrometer is used 
in the pulsar binning mode wherein each correlation function was recording into 
one of $2^N$ (e.g. 32) pulse phase bins that resulted in $2^N$ phase binned spectra covering
a radio frequency bandwidth (e.g. 4 MHz for Parkes telescope, subdivided
into 2048 
spectral channels typically $\sim 2 \rm kHz \; \rm or \sim 0.4 \rm \; km \; s^{-1}$ wide).
The spectra obtained in the phase bins that had the pulsar pulse were collapsed into a single ``pulsar-on"
spectrum. Similarly, the spectra recorded for
off-pulse phase bins were integrated into a single ``pulsar-off" spectrum.
The so-called {\it pulsar spectrum} was formed from the difference of
{\it pulsar-on} and {\it pulsar-off} spectra and normalized by the mean pulsar
flux. 
A frequency switching that takes the central radio frequency away from
the line was used to flatten the baseline.
This method of constructing the {\it pulsar spectrum} 
removes any in-beam molecular or atomic line {\it emission}
due to the broad telescope beam and measures the pulsar signal alone absorbed by
any intervening OH in intrabinary or interstellar medium. 
\smallskip

We estimate the typical exposure requirements for Green Bank Telescope (GBT)
if it were to be used for OH line detection from pulsars similar to those listed
in Table \ref{obstab}. 
%
We note that all pulsars listed in Table \ref{obstab} have
typical {\it mean} S1400 flux densities of 3-4 mJy (except the one
in globular cluster NGC6544, namely PSR J1807-2459A). 
These pulsars also have a $10\%$ duty cycle, which implies that the
peak flux during the pulse on phases will be approximately 10 times
higher, although the pulse on phase is only 1/10th the pulsar full period.
The radiometer equation gives the sensitivity of a radio telescope and receiver
system to pulsed signals in terms of a 
threshold
flux density $S_{mean}$ \citep{2012hpa..book.....L}:
\begin{equation}\label{eqn:signal-to-noise-cont}
S_{mean} = \beta \frac{(S/N)_{mean} T_{sys}}{G \sqrt{n_p t_{int}\Delta f}}\sqrt{\frac{W}{P-W}}, 
\end{equation}
where $T_{sys}$ is the system noise temperature (including sky and receiver noise),
$G$ the telescope gain to convert between temperature and flux density (in the Rayleigh Jeans limit),
$n_p$ is the number of orthogonal polarizations summed in the signal, $t$ is the integration time
for the pulsar observation, $\Delta f$ is the observing bandwidth, $(S/N)_{mean}$ is the threshold
signal to noise for detection, 
$W$ is the `pulse on' duration, $P$ is the
pulse period (rotational period of the neutron star), and $\beta \simeq 1$ denotes a factor of 
imperfections due to
the
digitization of the signal in the receiver system and other effects.

For a pulsar with a mean flux $3 \; \rm mJy$ and a signal to noise ratio
$(S/N)_{mean} \sim 9$ (see discussion after Eq. \eqref{eqn:eq-width-final}),
observed with
GBT\footnote{
The parameters of the GBT L-band Receiver listed in Eq. (\ref{exposuretime}) 
can be found in Table 3 and Fig 3
of the GBT Proposers Guide at: https://science.nrao.edu/facilities/gbt/proposing/GBTp. The actual time 
reported on the left hand side of Eq. (\ref{exposuretime}) has
been calculated using the Sensitivity Calculator for VEGAS spectrometer at: 
https://dss.gb.nrao.edu/calculator-ui/war/Calculator$\_$ui.html. This is based on the assumption
that the typical duty cycle has $W/P \sim 0.1$. 
Calculated exposure time depends upon several factors, e.g. whether signal and reference observations
are differenced, the ratio of observing time on signal and reference, etc.
The time reported in Eq. (\ref{exposuretime}) 
does not difference signal and reference
observations. Continuum flux is also measured in "signal" observations. 
}, 
Eq. (\ref{eqn:signal-to-noise-cont}) 
can be rewritten as:
\begin{align}\label{exposuretime}
\frac{t_{int}}{5 \; \rm hr} 
= \Big(\frac{T_{sys}}{16 \; \rm K}\Big)^2\Big(\frac{G}{2 \; \rm K/Jy}\Big)^{-2}\Big(\frac{S_{mean}}{0.2 \; \rm mJy}\Big)^{-2}\Big(\frac{W/P}{0.1}\Big) \nonumber \\
\times  \Big[\Big(\frac{n_p}{2}\Big)\Big(\frac{\Delta f}{0.4 \; \rm km/s}\Big) \Big]^{-1} 
\end{align}
The minimum detectable flux for a given pulsar (with $P$ and $W$)
is usually set by the threshold signal to noise ratio $(S/N)$. 
But since we are interested in absorption line profile in the
{\it pulsar spectrum}, which flux are we referring to? 
If we set $S_{mean}$
to be the continuum flux of the {\it pulsar spectrum} outside the channels occupied
by the absorption line, the flux there is relatively high. However, what matters
for line profile determination is that our flux uncertainty should be low enough 
to measure the {\it residual} 
between the continuum $S_c$ (assumed to be constant over a 
narrow line width say of Gaussian profile\footnote{
In terms of the spectrometer bandwidth, a Gaussian profile with line broadening $\xi_m$
due to microturbulence and thermal broadening has
$FWHM = 1.6652 (\nu_0/c) \sqrt{(2 kT/m + \xi_m^2)}$.
}) and the flux density at the line center (where the 
optical depth is $\tau_0$) 
with 
a
high enough significance.
Since the absorption of pulsed radiation in the OH band
is variable with time due to
the
pulsar orbital motion, the 
determination of its magnitude can be limited by
the
errors arising out of short exposure times.
The equivalent width 
for the absorption line itself (for a Gaussian line) can be written as:
$$EW = 1.06 \; \tau_0 \; FWHM $$
The equivalent width can be written in terms of the
flux density in the continuum $F_c(\lambda)$ and in the line $F(\lambda)$
in a wavelength interval $\lambda_1 < \lambda < \lambda_2$ (outside this regime 
$F(\lambda) = F_c(\lambda)$) as:
$$EW_{\lambda} = \Delta \lambda - \int_{\lambda_1}^{\lambda_2} \frac{F(\lambda)}{F_c(\lambda)} d\lambda $$
where $\Delta \lambda = \lambda_2 - \lambda_1$. By applying the mean value theorem this
can be re-expressed as \citep{2006AN....327..862V}:
$$ EW_{\lambda} = \Delta \lambda \Big[ 1 - \frac{\overline{F(\lambda)}}{\overline{F_c(\lambda)}}\Big] $$
where $F_c(\lambda)$ is measured outside the line
region and interpolated across the line where the
absorption is taking place and the overbars denote the average value
of a variable. Both $F(\lambda)$ and $F_c(\lambda)$ have statistical errors.
If their errors 
are not correlated, then their standard
deviations can be determined separately. 
On the other hand the error estimate of the equivalent width of a spectral line
has been derived by \citet{2006AN....327..862V}. The overall uncertainty of the EW
is determined by two factors: the photometric uncertainty of the system and
the uncertainty of the continuum estimation over the line.  
For low absorption line fluxes \citet{2006AN....327..862V} show that the standard error of the 
equivalent width $\sigma (EW_{\lambda})$ can be written as:
$$ \sigma (EW_{\lambda}) = \frac{(\Delta \lambda - EW_{\lambda})}{(S/N)} \times \Big[1+ \frac{\overline{F_c}}{\overline{F}}\Big]^{1/2}$$
where the relevant $S/N$ ratio is the $S/N$ in the undisturbed continuum and $\Delta \lambda$ is a 
measure of the line width in wavelength units. For weak lines, i.e. where the depth of the line
is very small, $\overline{F} \sim \overline{F_c}$ and the above relation reduces to:
$$\sigma (EW_{\lambda}) =  \sqrt{2} \frac{(\Delta \lambda - EW_{\lambda})}{(S/N)}$$
%
This then leads to, using $\Delta \lambda - EW_{\lambda} = \Delta \lambda (\overline{F}/ \overline{F_c})$,
%
\begin{equation}\label{eqn:eq-width-significance}
 \frac{EW_{\nu}}{\sigma(EW_{\nu})}
= (S/N) \frac{EW_{\nu}}{\Delta\nu}\times\Big[\frac{\overline{F}}{\overline{F_c}} \big(1 + \frac{\overline{F}}{\overline{F_c}}\big) \Big]^{-1/2}
\end{equation}
where $\Delta \nu = (c/\lambda^2) (\lambda_2 - \lambda_1)$.
Here 
$(S/N)$ $(= (S/N)_{mean}$) is
determined by
the
observational and telescope parameters
as in Eq. \eqref{eqn:signal-to-noise-cont}. 
For weak or low absorption line strengths, one has $\overline{F} \sim \overline{F_c}$, and the flux 
dependent factor in square brackets in Eq. \eqref{eqn:eq-width-significance}  
is about $1/\sqrt{2}$. Similarly the ratio of FWHM to $\Delta \nu$ that arises out of the ratio 
$(EW_{\nu}/\Delta \nu)$ is also of the order of unity.
Therefore given a $\tau_0$ the significance of the line detection in terms of its equivalent width
is:
\begin{equation}\label{eqn:eq-width-final}
 \frac{EW_{\lambda}}{\sigma(EW_{\lambda})}
 \approx  0.3 \Big(\frac{\tau_0}{0.3}\Big) (S/N)_{mean} \\
\end{equation}

Against the very bright pulsar-on phase, it will be relatively easier
to measure 
the
changes in 
the
flux density in the OH-line profile, since the 
high flux density at a specific frequency channel can be relatively easily
determined.
The local continuum of the {\it pulsar spectrum}
can be determined much more accurately than in the line absorption frequency channels, since many more
channels 
(compared to where the line is located) 
can be averaged together in the continuum during data processing to get a good $(S/N)_{mean}$.
If the absorption
in the 1667 MHz band has, for example, an optical depth at the line center
$\tau_0 = 0.3$ then the 
significance of the EW measurement would be 
lower than that of pulse detection in the continuum part of the {\it pulsar spectrum}.
With 
increasing $\tau_0$, the equivalent width will be determined at a significance $(S/N)_{mean}$ approaching the
pulsar signal detection itself (in the continuum).
Thus to effect a detection of the absorption line with a similar $\tau_0$ (and with $(EW/\Delta EW) = 3$),
one requires
a significance of detection of the {\it pulsar spectrum}
of $(S/N)_{mean} \sim 9$
while for deeper absorption lines (with higher $\tau_0$), a significant detection of the OH absorption
line may be effected at even a lower 
$(S/N)_{mean} \sim 3-5$.
Our estimate of
the
telescope resource requirements before (e.g. Eq. \eqref{exposuretime}) however is 
based on the conservative 
case of moderate absorption depths and line-widths.

The dip in the flux density would decrease towards
the absorption line wings over a velocity range whose
magnitude could be similar or somewhat larger than
the thermal speed
(about $0.4 \; \rm km \; s^{-1}$) of the OH gas,
i.e. about $ 2-3 \; \rm km \; s^{-1}$ (we note that \citet{2008ApJ...677..373M} gives
the FWHM for PSR B1718-35 OH absorption in the {\it interstellar medium}
to be about $3\; \rm km \; s^{-1}$ at 1667 MHz; in the binary system, the wind speed from the companion and
the OH line width
is likely to be larger). 
The orbital velocity
scale is typically several $ 100 \; \rm km \; s^{-1}$ 
for a system like PSR B1957+20, although because of the
mainly tangential 
swept back wind, the 
velocity width in the 
radiatively connected region
along the line of sight
is likely to be
substantially 
smaller. Thus, the OH
absorption at orbital phases near the eclipse may be distributed over a velocity width 
$\leq 10 \rm km s^{-1}$.
Note that since only times close to the total eclipse would have
substantial OH absorption, only that fraction of the orbital period
(e.g. the $\sim 50^m$ periods around ingress and egress for PSR B1957+20)
are relevant and the required total exposure have to be distributed over these
epochs. Since these periods last together about 10-15\% of the full orbital phase
(for the case of PSR B1957+20), the total exposure time has to be spread
over many orbital cycles.
Finally, it will be efficient to use a sufficiently large bandwidth of the spectrometer
in several spectral windows to cover all L-band OH lines (1612-1720 MHz) and HI line
simultaneously.

Thus, each pulsar listed in Table 1 (first four rows)
requires about 5 hrs of telescope time with the typical
sensitivity of 
GBT.
%
Pulsars in the declination range $0-40\degr$ 
are accessible to Arecibo, which is most sensitive existing telescope.
In fact, GBT, Arecibo, and
Parkes Radio Telescope each detected one of the pulsars listed in Table \ref{OH-absorption-detections}.
Future telescopes like the Square Kilometer Array (SKA) will have typically
ten (SKA1-Mid) to hundred times (full SKA) flux sensitivity compared to GBT.
With SKA-1 increased sensitivity in L-band (Band3) using low noise
amplifiers, it will be possible to carry out such molecular
line detections down to a pulsar mean flux density level
of $0.7\; \rm mJy$ or lower, opening up several other known pulsar targets
for similar studies.
In addition, many newly discovered pulsar targets
will become available for studies of composition of the winds from the
companion. 

\subsection{Search for gas in a binary pulsar system through gated spectral line interferometry} 
\smallskip
So far our discussion of absorption spectroscopy in the {\it pulsar spectrum} has been in the
time (pulsar rotational phase) vs radio frequency domain and in the context of single dish telescopes. 
In section 4.1 we have discussed that detection of 
orbital phase modulation can distinguish the intrabinary atomic/molecular gas 
from the circumbinary gas or that in the interstellar medium.
By employing multiple antenna dishes in an interferometer, it is also possible to add
a third domain, namely spatial information in the plane of the sky. While it is not
possible at present to spatially resolve a binary system at the relevant distances of pulsars,
any detection of molecular or atomic gas (through absorption or stimulated emission)
coincident with the pulsar position can be expected to be associated with the
binary system if it is point like and spatially coincident. 
This is especially so
if the pulsar
occurs out of the galactic plane where the occurrence of such gas clouds is 
rarer in the interstellar medium. 
The primary advantage offered by interferometric line absorption studies is the ability to resolve out the
foreground neutral atomic/molecular line emission and thus yield an uncontaminated measure of the
absorption line profile that traces gas in the narrow beam subtended by the background source
(in this case the pulse-on phase of the pulsar radiation). Moreover, it is possible with
interferometric systems to achieve high spectral dynamic range, which opens up the possibility
of measuring small optical depths, which can probe a Warm Neutral Medium
(WNM) \citep{2013MNRAS.436.2352R}.
\citet{2007ASPC..365...28W} showed that for PSR B1849+00 whose line of sight
passes near the edge of the SNR Kes 79 had far smaller optical depth for OH absorption
in the pulsar-off phases than in the pulsar-on OH spectra ($\tau (1667 MHz) = 0.02 \; \rm vs \; 0.9$),
which they explained as the pulsar-Earth pencil beam of radiation
having intercepted dense ($> 10^5 \; \rm cm^{-3}$) and small
(typical angular size $<15"$) cloudlets in the pulsar ISM spectra. In the pulsar-off spectrum,
the sampling of absorption is across the full telescope beam that represents a larger solid
angle average across a clumpy ISM consisting of high optical depth, small molecular
cloudlets embedded in lower density medium. 

As is well known, the significance of detection of a pulsating point source can be largely improved
by removing the off-pulse noise. A background sky subtraction procedure for the pulse on - pulse off ``gated" 
image constructed by an interferometer (such as the ATCA, \citep{2000ApJ...541..367C}) 
can unambiguously identify the location of the pulsed emission. 
Millisecond pulsars (MSPs, discovered by {\it Fermi} with poor spatial information)
which require high time resolution for the gating procedure have been localized
to an accuracy of $\pm 1^"$ in the {\it on - off} gated image plane at GMRT.
GMRT observations
use 
a coherently de-dispersed gating correlator of the multiple antenna outputs
that accounts for, at the same time, orbital motions of the MSPs while interferometer visibilities
are folded with a topocentric rotational model derived from periodicity search simultaneously
with the beamformer output
\citep{2013ApJ...765L..45R}. 
The positional accuracy of even the faint MSPs obtained from the {\it on-off}
gated image improve with the S/N of the pulsar detection. 
Even though at 322 MHz, the GMRT synthesized 
beam has a positional accuracy of FWHM $10^"$, an accuracy of
$\pm 1^"$ has been obtained for the timing position of the pulsar with a S/N of 5,
and the astrometric accuracy accelerates the
convergence in pulsar timing models 
for newly discovered pulsars such as the {\it Fermi} MSPs \citep{2013ApJ...765L..45R}.
This reduces the telescope time requirements for subsequent follow-up timing observations
and reduces the effect of large covariances in the timing fit between pulsar
position and spin period derivative, $\dot P$. 
While GMRT cannot observe in the OH bands, it can do so in the 
HI (1420 MHz) band.
For example, GMRT interferometric observations of Galactic HI 21 cm absorption spectroscopy towards 32 
compact bright extragalactic radio sources (with L-band flux greater than 3 Jy) 
have led to the detection of
Warm Neutral Medium of
spin temperature of several thousand degrees K \citep{2013MNRAS.436.2352R}.
In the HI band these
observations achieved typically a
velocity resolution of $\sim 0.4 \; \rm km \; s^{-1}$
and 
a velocity coverage of $\sim 105 \; \rm km \; s^{-1}$
by using a single IF band with two polarizations and a baseband bandwidth of 0.5 MHz
subdivided into 256 channels.

At the VLA, THOR - The HI, OH, Recombination
Line Survey of the Milky Way has been undertaken to study atomic, molecular and 
ionized emission of Giant Molecular Clouds (GMC) in our galaxy \citep{2015A&A...580A.112B}.
These observations are interferometric (e.g. the Pilot survey for HI from GMC W43 was in the C-array
of VLA) and have high spectral resolution (channel width of 1.953 kHz, with 
$\Delta v \sim 0.41 \; \rm km \; s^{-1}$ in the HI 21cm line, while that for OH lines for the
region around the same target the channel spacing was $0.73 \; \rm km \; s^{-1}$ 
at 1612 MHz \citep{2016MNRAS.455.3494W}).
The synthesized beam size was of the order of $20^" \times 20^"$
but the absolute positional accuracy of a point source with a $5\sigma$ detection would
be: $\Theta_{Beam}/2\sigma = 2^"$. Thus, the VLA C-array and GMRT
would achieve similar spatial resolutions for OH/HI bands centered around 
a target binary MSP at about 1 kpc, with $\sim 0.01 \; \rm pc$ spatial scale.
Since spatial densities of molecular or HI clouds along a narrowly defined line of sight
towards high galactic latitude are likely to be small, a detection of absorbing gas
is likely to be physically associated with the pulsar system itself, especially
if the column densities are high. 

The typical rms noise in THOR Pilot survey was 19 mJy/beam for the OH lines
and 9 mJy/beam for the HI band which was achieved in 8 min pointing of the VLA.
The survey has not exploited any underlying time dependent structure (e.g.
a pulsar) of the spectral line signal. As we argue in section 5.1, for a pulsar of mean flux density
of 3 mJy being obscured by companion wind of optical depth $\tau \sim 0.3$, 
a typical $10\sigma$ signal for the pulsar continuum required for adequate line detection
would lead to $1\sigma$ rms of 0.3 mJy. If there were no pulsed signal, such a low rms could be
achieved with the VLA if we scale the THOR exposure ($8 \; \rm min$) 
by the ratio of the square of the respective
flux rms, to obtain an overall exposure of $(19 \; \rm mJy/0.3 \; \rm mJy)^2 \times 8 \; \rm min = 
534 \; \rm hr$. 
However, for a MSP with a duty cycle that is $10\%$ of its spin period, if we can (phase)gate the
pulsar signal and construct the {\it pulsar on - off spectra} as described before
we will gain a factor of
$1/10$, thus reducing the exposure requirement to about $53 \; \rm hr$. This is still too large
for a single pulsar target. 
While such a capability may not exist at present for the OH band,
the HI line transitions may be more promising, requiring about $12 \; \rm hr$ of exposure (with similar 
scaling and assumptions
of optical depth in the HI line absorption), provided adequate hardware and computational resources are
available.

\section{Discussion}

Molecular line detection would indicate
the presence of constituent elements in the vicinity
of the planet or the ultra-low mass companion.
An Oxygen rich atmosphere
could indicate a
CO degenerate core of a helium burnt star as the companion.
A simultaneous detection of neutral hydrogen
and OH lines would indicate whether the gas has solar metallicity
or has metals which are substantially super-solar, as might be
the case if the companion's external surface which may have
once contained substantial fraction of hydrogen is stripped off of its
external layers.
Detection of OH lines or the 
atomic hyperfine transition line of $^3He^+$ ($^2S_{1/2}, F = 0-1$)
at 8.665 GHz would lead to constraints on the present state, e.g.
the mass and radius of the companion.

The OH molecule is observed in a variety of astrophysical
systems, including comets and the ISM. OH is usually
formed by the dissociation of $\rm H_2 O$ molecule, so it is often considered
a proxy for water. The rotational ground state of the OH molecule 
has four transitions: two ``main" lines at 1667 MHz and 1665 MHz (with
the largest Einstein A coefficients, e.g. $7.7 \times 10^{-11} \; \rm s^{-1}$)
and two ``satellite" ones at 1720 MHz and 1612 MHz 
\citep{1996tra..book.....R}. The pumping mechanisms
differ from one astrophysical system to another. In the case of
comets, heating of their nuclei leads to evaporation of water and photo-dissociation
due to solar radiation and the level population distribution is
dominated by UV pumping by solar radiation. Absorption features in the
solar spectrum can in some circumstances cause population inversion of
the ground state. 
The 1665 and 1667 MHz main lines are often associated with star forming
regions whereas the 1612 MHz lines are associated with evolved stars.
All these lines are due to inversions arising through pumping of far-infra
red photons. In contrast, the OH (1720 MHz) masers are pumped through
collisions and far-IR radiation effectively causes the destruction of the
population inversion required for maser action.
A strong maser inversion for the OH (1720 MHz) line is collisionally
excited at temperatures 30-120 K 
\citep{1999ApJ...511..235L,1999ApJ...525L.101W}.
If maser emission at 1720 MHz and
absorption at conjugate frequency 
of 1612 MHz are detected, together these would constrain 
upper and lower limits of $(1\times 10^{14} \; \rm cm^{-2}) \rm (km \; s^{-1})^{-1} < N_{OH}/ \Delta \rm v < (1\times 10^{15} \; \rm cm^{-2}) \rm (km \; s^{-1})^{-1}$ 
when the gas column
becomes thick to far-infra-red pump photons. When 
column densities higher than  $(1\times 10^{15} \; \rm cm^{-2}) \rm (km \; s^{-1})^{-1}$
are encountered this would lead to absorption in the 1720 MHz line \citep{2005Sci...309..106W}.
Maser amplification requires large column densities of OH molecules
($10^{16} - 10^{17} \; \rm cm^{-2}$) with small velocity gradients.
\citet{2011MmSAI..82..703F} has argued that such masers can occur
preferentially where the observer's line of sight velocity gradient
is small, as in edge on geometries of transverse compressional shocks
in supernova remnants interacting with molecular clouds.
The geometry of the pulsar driven swept back winds from the companions 
(as for example indicated by the pattern of dispersion measure related
delays in the black widow pulsar PSR B1957+20) would
also offer favorable geometry of small velocity gradients along the
line of sight, especially in cases where the orbital speed of the
companion is large.
Therefore the detection of satellite lines in emission or
absorption may probe the column density (velocity) gradients
in the intrabinary region.  
Moreover, the search for OH absorption and possible detection of
stimulated emission on timescales corresponding to duty cycles of millisecond pulsars
will probe much shorter timescales of maser action by 2-3 orders of magnitude than known so far 
and may constrain the aspect ratios of the filamentary
structures far beyond what are currently thought to be
responsible for maser amplification in circumstellar or interstellar medium \citep{eli91}.
Black widow and red back pulsars have the characteristics
that make them good targets for OH line detection observations
since pulsar timing and optical observations allow us to determine
their geometry, including the eclipsing region, well and
their short orbital periods allow for repeated observations over many 
orbits within reasonable timelines
to build up gated exposure time on the radio pulses near eclipse
ingress and egress.

\section{Acknowledgements}

We thank Mark Reid and Jonathan Grindlay
for their comments on the manuscript. A.R. thanks Vicky Kaspi,
Avinash Deshpande
Sayan Chakraborti and Wlodek Kluzniak for discussions.
He thanks the Fulbright Foundation for a Fulbright-Nehru
Fellowship 
during a sabbatical leave
from Tata Institute of Fundamental
Research and thanks
the Director and staff of
the Institute for Theory and Computation,
Harvard University 
for their hospitality during his visit.
We acknowledge the use of
Australia Telescope National Facility (ATNF) Pulsar
Catalogue and Paolo Freire's website on
"Pulsars in Globular Clusters and the NASA Astrophysics
Data System (ADS) search engines.
We thank Nirupam Roy for discussions
on spectral line interferometry
for HI and OH lines and for pointing out
the THOR pilot papers.
We thank 
the referee 
Joel Weisberg
for many insightful comments
which helped us
to improve the paper. 
We thank 
Thomas Tauris for his correspondence on the manuscript.

\bibliographystyle{apj}

\end{document}